\documentclass[journal]{IEEEtran}
\IEEEoverridecommandlockouts
\makeatletter
\def\endthebibliography{%
	\def\@noitemerr{\@latex@warning{Empty `thebibliography' environment}}%
	\endlist
}
\makeatother

\usepackage{cite}
\usepackage{amsthm,amssymb,amsfonts}
\usepackage{amsmath}
\usepackage{graphicx}
\usepackage{subfig}
\usepackage{caption}
\usepackage{xcolor}
\usepackage{array}
\usepackage{verbatim}
\usepackage{multirow}
\usepackage{booktabs}
\usepackage{makecell}
\usepackage{balance}
\usepackage{floatrow}
\usepackage{subfig}

\usepackage{enumerate}
\usepackage{enumitem}
\usepackage{caption}
\usepackage{pifont}
\definecolor{mymag}{RGB}{255, 0, 255}
\definecolor{mygreen}{RGB}{0, 176, 80}

\PassOptionsToPackage{hyphens}{url}

\usepackage{hyperref}

\usepackage{tikz}

\def\ie{\textit{i.e.}}

\def\eg{\textit{e.g.}}

\begin{document}
	
	\title{Wireless Powering Internet of Things with UAVs: Challenges and Opportunities}

	\author{Yalin Liu, Hong-Ning Dai,~\IEEEmembership{Senior Member,~IEEE,} 
		Qubeijian Wang,
		Muhammad Imran,
		Nadra Guizani
		
		\thanks{Y. Liu, is with Faculty of Information Technology, Macau University of Science and Technology, Macau. e-mail: yalin\_liu@foxmail.com.}
		
		\thanks{H.-N. Dai is with Department of Computing and Decision Sciences, Lingnan University, Hong Kong. email: hndai@ieee.org.}%
		
		\thanks{Q. Wang is with School of Cyber Security, Northwestern Polytechnical University, China. email: qubeijian.wang@nwpu.edu.cn.}%
		
		\thanks{M. Imran is with the School of Engineering, Information Technology \& Physical Sciences, Federation University Australia, Brisbane, Australia. email: dr.m.imran@ieee.org.}%
		
		\thanks{N. Guizani is with the School of Electrical and Computer Engineering at Washington State University, U.S.A. email: nadra.guizani@wsu.edu.}%
		
	}

	\maketitle
	
	\begin{abstract}
		Unmanned aerial vehicles (UAVs) have the potential to overcome the deployment constraint of Internet of Things (IoT) in remote or rural area. Wirelessly powered communications (WPC) can address the battery limitation of IoT devices through transferring wireless power to IoT devices. The integration of UAVs and WPC, namely UAV-enabled Wireless Powering IoT (Ue-WPIoT) can greatly extend the IoT applications from cities to remote or rural areas. In this article, we present a state-of-the-art overview of Ue-WPIoT by first illustrating the working flow of Ue-WPIoT and discussing the challenges. We then introduce the enabling technologies in realizing Ue-WPIoT. Simulation results validate the effectiveness of the enabling technologies in Ue-WPIoT. We finally outline the future directions and open issues.
	\end{abstract}
	

	\IEEEpeerreviewmaketitle

	\section{Introduction}
	
	We have witnessed the proliferation of IoT, which has been widely adopted in diverse urban applications, such as smart home, smart healthcare, smart industry, smart grid. IoT has been typically deployed in the scenarios with the availability of communications infrastructure, such as base stations, access points and IoT gateways. 
	However, the deployment and maintenance of infrastructure nodes inevitably bring huge operational expenditure. Moreover, it is difficult to deploy wireless infrastructure nodes at remote or rural area (\eg, forest surveillance and livestock monitoring). In addition, IoT devices are also life-limited due to their built-in battery limitations. These two fundamental constraints of IoT prevent its wide deployment in remote scenarios.
	
	The recent advances in UAVs bring opportunities to overcome the limitations of IoT. Related work has regard UAVs as elastic communication nodes to increase the communication coverage and enhance the network capacity~\cite{Foto:CST19}.	Particularly, UAVs have the advantage in aerial mobility in contrast to ground vehicles. For example, UAVs can be flexibly dispatched to remote, rural or disaster area where ground vehicles can hardly or cannot directly reach~\cite{Foto:CST19}. Hence, UAVs can potentially foster remote IoT applications such as environment monitoring, farm observation and emergency communications. 
	
	Radio frequency (RF)-based Wireless Power Transfer (WPT) is a promising method to prolong the life of the battery-limited nodes. In previous literature, RF-WPT has been presented to support wireless communications. The integration of RF-WPT and wireless communications leads to a new type of wireless communications called WPC~\cite{huang2015cutting}. Particularly, WPC has two main application scenarios: WPC networks and simultaneous wireless information and power transfer (SWIPT). In WPC networks, IoT nodes first harvest wireless energy, which is then used for data transmission. SWIPT aiming to achieve WPT and information transmission simultaneously in the same channel has a critical hardware requirement on IoT nodes. Therefore, WPC is more preferred for energy-limited IoT devices, thereby overcoming the second constraint of IoT.

\begin{figure*}
\ffigbox[16.2cm]{%
\begin{subfloatrow}
  \ffigbox[\FBwidth][]
    {\caption{Applications of Ue-WPIoT}\label{sfig:applications}}
    {\includegraphics[height=8cm]{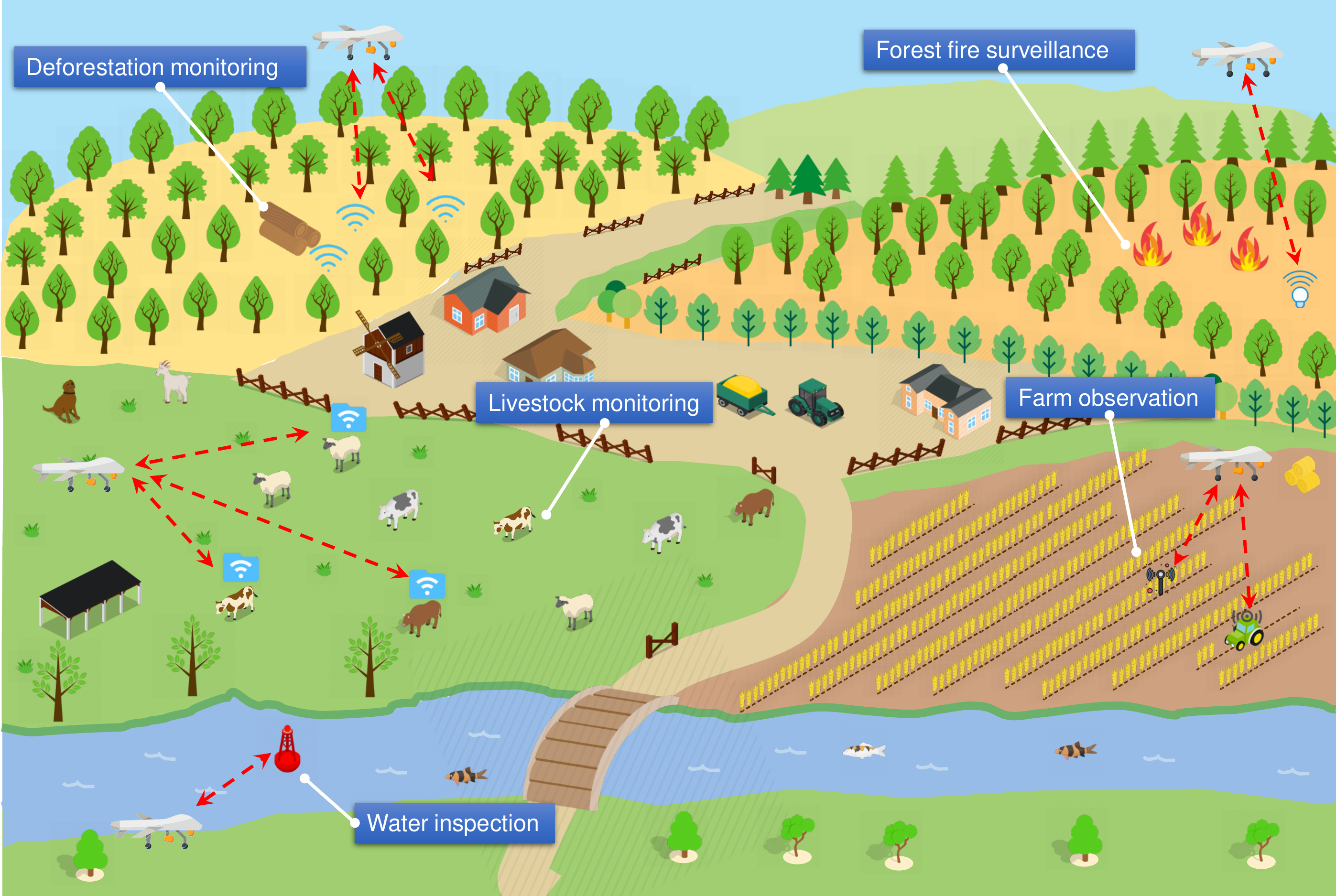}}
\end{subfloatrow}
\hspace*{\columnsep}
\begin{subfloatrow}
  \hsize0.2\hsize
   \vbox to 8.55cm{
  \ffigbox[\FBwidth]
    {\caption{Wake-up procedure}\label{sfig:process1}}
    {\includegraphics[height=3.6cm]{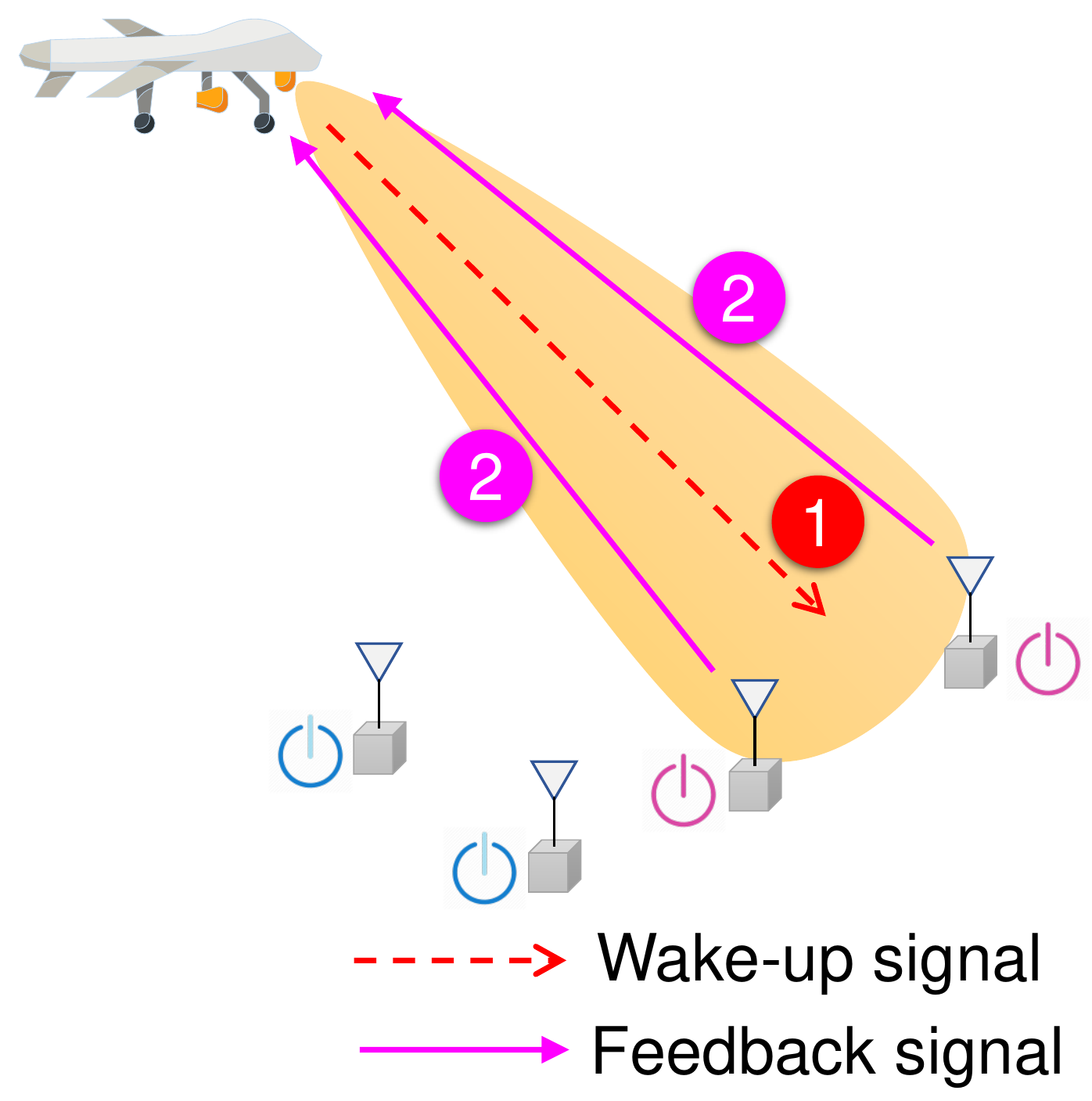}}
    \vss
  \ffigbox[\FBwidth]
    {\caption{Wireless powering and data transmission}\label{sfig:process2}}
    {\includegraphics[width=3.6cm]{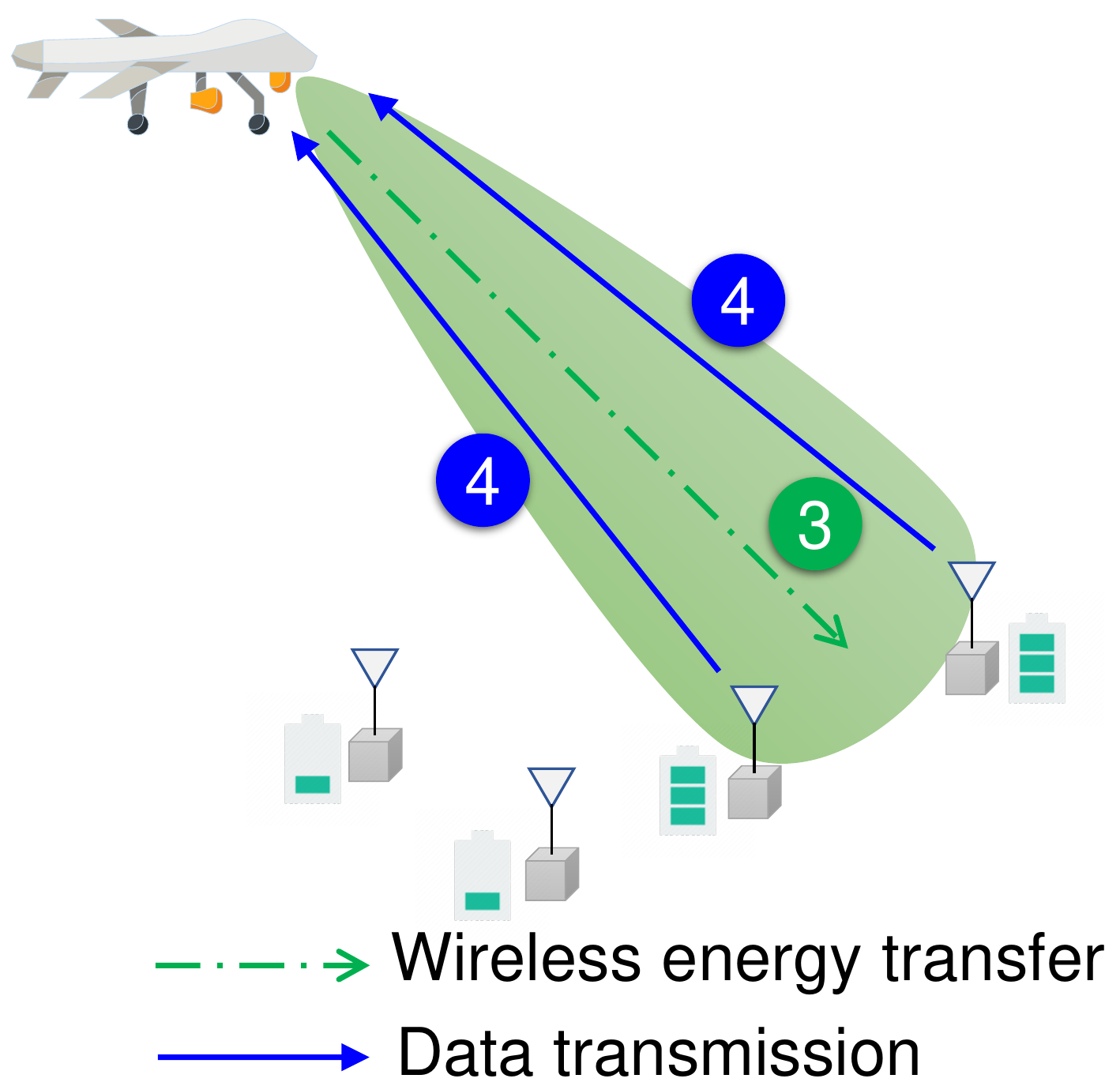}}
   }
\end{subfloatrow}
}
{\caption{An overview of Ue-WPIoT}\label{fig:system}}
\end{figure*}

	The integration of UAVs with WPC brings opportunities to extend the IoT applications from cities to remote or rural areas. Combining with WPC, a UAV has the capability to supply energy to IoT nodes for their data transmission. We name such UAV-enabled wireless powering Internet of Things as Ue-WPIoT.  
	However, the realization of Ue-WPIoT suffers a number of challenges including the limited powering range, energy efficiency optimization, flying trajectory design. This article aims at investigating solutions to those challenges when realizing Ue-WPIoT. As a summary, this article has the following contributions.
	
	We present Ue-WPIoT to support remote IoT applications (in Section~\ref{sec:system}). In Ue-WPIoT, IoT nodes that sleep (or hibernate) to save the energy can be activated by the wake-up signals emitted from a UAV. Consequently, the UAV can locate the IoT nodes and then transmit wireless energy to IoT nodes. Thereafter, IoT nodes can transmit the collected data to the UAV via using the harvested energy. Ue-WPIoT can essentially enable a plethora of applications including forest deforestation monitoring, livestock monitoring, water inspection, farm observation and forest fire surveillance as shown in Fig.~\ref{fig:system}. 
	
	We then discuss the challenges before the realization of Ue-WPIoT. Firstly, the RF-based WPT usually experiences high attenuation over a distance, thereby limiting the effective communication range. Secondly, the overall energy efficiency during WPT and data transmission needs to be optimized. Thirdly, the trajectory design of UAVs needs to be optimized after the joint consideration of multi-node communications and the travelling path length together.

	We next introduce enabling technologies to address the challenges in realizing Ue-WPIoT (in Section~\ref{sec:technology}). Specifically, we adopt the adaptive energy beam-forming technologies in the UAV, thereby extending the powering distance. Moreover, we formulate a common optimization model to optimize the energy efficiency in Ue-WPIoT. Finally, we design the optimal flying trajectory of the UAV. The numerical results also validate the effectiveness of the promising technologies (in Section~\ref{sec:simu}). We also outline the future directions in Ue-WPIoT (in Section~\ref{sec:conc}).

	\section{UAV-enabled Wireless Powering IoT}
	\label{sec:system}
	
	\subsection{Working flow}
	\label{subsec:system-arch}	
	Fig.~\ref{fig:system} presents the applications of Ue-WPIoT and the working procedure of Ue-WPIoT. As shown in Fig.~\ref{sfig:applications}, we mainly consider a fixed-wing UAV with sufficient battery capacity or fuels to support a long flight to remote or rural area. Ue-WPIoT can support a diversity of applications including deforestation monitoring, livestock monitoring, water inspection, farm observation and forest fire surveillance. In these scenarios, a UAV can first transfer wireless energy to charge IoT nodes which send back the collected data to the UAV. One UAV-IoT communication procedure consists of two stages as shown in Fig.~\ref{sfig:process1} and Fig.~\ref{sfig:process2}. 
	
	
	We next describe the working flow of one Ue-WPIoT communication as follows: 1) \textit{Wake-up procedure.} Since IoT nodes have usually been in sleeping or hibernating mode to save energy, the UAV wakes up IoT nodes by transmitting the Wake-up Radio (WuR) signals~\cite{7990121} (\ie, Step \scalebox{1.03}{\textcolor{red}{\ding{182}}} as shown in Fig.~\ref{sfig:process1}). Then IoT nodes can be activated when they detect enough power from the wake-up radio signals. The activated IoT nodes transmit the feedback signals to the UAV which can then identify the activated IoT nodes (\ie, Step \scalebox{1.03}{\textcolor{mymag}{\ding{183}}} as shown in Fig.~\ref{sfig:process1}). In the Ue-WPIoT system, each UAV is equipped with an antenna array, which can effectively differentiate arrivals of multiple feedback signals and locate the precise orientation of each IoT node~\cite{6568923}. 2) \textit{Wireless powering and data transmission.} The UAV next transfers wireless energy toward the activated IoT nodes. With an antenna array, the UAV is capable of generating a sharp beam toward IoT nodes, thereby improving the wireless powering efficiency of IoT nodes~\cite{alsaba2018beamforming} (\ie, Step \scalebox{1.03}{\textcolor{mygreen}{\ding{184}}} as shown in Fig.~\ref{sfig:process2}). IoT nodes can leverage the harvested energy to transmit the data to the UAV (\ie, Step \scalebox{1.03}{\textcolor{blue}{\ding{185}}} as shown in Fig.~\ref{sfig:process2}).
	
	During the above procedure, Steps \scalebox{1.03}{\textcolor{red}{\ding{182}}} and \scalebox{1.03}{\textcolor{mymag}{\ding{183}}} can activate and then locate the IoT nodes, thereby paving the way for the following wireless power transfer and data transmission. It is worth mentioning that the performance of data transmission greatly relies on the harvested energy in the previous wireless powering step. Thereafter, IoT nodes can harvest enough energy from the UAV (\ie, Step \scalebox{1.03}{\textcolor{mygreen}{\ding{184}}}) to support the data transmission in Step \scalebox{1.03}{\textcolor{blue}{\ding{185}}}. Ue-WPIoT can essentially support multiple concurrent communications since multiple nodes can be activated as the same time when they are close to each other. In this case, a multi-access mechanism (\eg, time-division-multiple-access or spatial division-multiple-access) shall be adopted to support multiple concurrent communications with different IoT nodes. 
	
	\subsection{Design Challenges}
	\label{subsec:challenge}
	
	Ue-WPIoT can promote the wide adoption of UAVs to connect IoT nodes in remote or rural area while the realization of Ue-WPIoT is faced with the following design challenges.
	
	\begin{itemize}
		\item \textit{Feasible Communication Range.} Generally, WPT has a much shorter range than data transmission as indicated in~\cite{Clerckx:JSAC19}. Thus, the feasible communication range of Ue-WPIoT is essentially limited by the achievable WET range. The communication range constraint may limit the wide application of Ue-WPIoT in different scenarios.  
		\item \textit{Energy-efficiency optimization.} UAVs are also suffering from the limited energy to support both flight propulsion and Ue-WPIoT communications. The energy consumption of the whole flight occupies the major proportion of the entire energy consumption. Even though UAVs can harvest energy from the ambience (\eg, solar panels), the ambience energy sources being susceptible to environment fluctuations cannot be the stable energy sources for UAVs. Thus, it is a critical issue to optimize the energy-efficiency with consideration of flight time, multi-access and trajectory design.
		\item \textit{Flying Trajectory Design.} In Ue-WPIoT, a UAV is expected to conduct tasks to visit multiple IoT nodes along a given trajectory. It is challenging to design the optimal trajectory, which requires the joint consideration of multiple factors, such as multi-node access, multi-task scheduling and the total flight time.  
		
	\end{itemize}
	
	\begin{table}[t]
		\centering
		\caption{\small Summary of Enabling Technologies to Address Ue-WPIoT Challenges}
		\renewcommand{\arraystretch}{1.5}
		\footnotesize
		\label{tab:solutions}
		\begin{tabular}{|p{3.8cm}|p{4.3cm}|}
			\hline
			\textbf{Challenges} & \textbf{Enabling Technologies}\\ 
			\hline\hline
			\multirow{2}{*}{Feasible Communication Range} & $\cdot$ Adaptive energy beam-forming \\
			 & $\cdot$ IoT EH Circuit Design \\
			\hline
			Energy-efficiency optimization & Resource allocation and optimization\\
			\hline
			Flying trajectory design & Joint trajectory optimization of multi-access and WPC\\ 
			\hline	
		\end{tabular}
	\end{table}
	
	\section{Enabling Technologies}
	\label{sec:technology}
	
	This section discusses several enabling technologies to address the above challenges of Ue-WPIoT. Table~\ref{tab:solutions} summarizes the state-of-the-art solutions in different aspects. 
	
	\subsection{Adaptive Energy Beamforming and IoT EH Circuit Design}
	\label{subsec:AE-BF}
	
	Ue-WPIoT is suffering from the limited communication range, which is mainly constrained by the WET range. The fundamental limitation of the WET range lies in much higher threshold for energy harvesting (EH) than that for information decoding~\cite{Clerckx:JSAC19}. The advent of energy beam-forming (BF) technology brings opportunities to overcome this limitation. In particular, the UAV employing a BF antenna can generate the BF energy toward a certain direction. The focused energy can significantly extend the powering distance.
	
	In Ue-WPIoT, the adoption of adaptive energy BF technology (AE-BF) can greatly improve the WET efficiency at IoT nodes. Specifically, a UAV with AE-BF capability can generate BF radio signal toward a certain orientation thereby improve the WET efficiency due to the focused energy. The adaptive BF is usually constructed via the prior-knowledge of channel state information (CSI), which can be obtained after analysing the feedback signals from the activated IoT nodes (\ie, Step \scalebox{1.03}{\textcolor{mymag}{\ding{183}}}). In addition, the adaptive BF vector can be optimized by maximizing the harvested power of the IoT nodes as in~\cite{alsaba2018beamforming}.
	

	With respect to the hardware design of IoT nodes, it may not be feasible to directly adopt the adaptive BF technology at IoT devices because it is difficult to equip expensive and bulky BF antennas at IoT nodes. However, appropriate configurations of IoT devices can extend the EH distance. For example, the sensitivity adjustment of the EH circuits can extend the powering distance~\cite{Clerckx:JSAC19}. Another design consideration is the radio frequency. In previous studies, most of prototypes of WuRs/energy harvesters are based on the frequency about 2.4GHz, 400MHz, and 900MHz, where 2.4GHz-based circuits are more compatible for 802.11-based wireless communication networks, while 400MHz or 900MHz based circuits may result in longer communication range.

	
	\subsection{Resource Allocation and Optimization}	
	\label{subsec:tech-reallocation}
	
	In Ue-WPIoT, practical powering process is susceptible to both the channel fluctuation and path loss effect since the IoT nodes can only transmit their data after harvesting enough wireless energy from the UAV. Meanwhile, different IoT nodes may gain a varied portion of wireless energy due to variations of air-to-ground (A2G) channels. Thus, it is necessary to design optimal resource allocation strategies for both UAVs and IoT nodes to optimize the energy efficiency of the entire Ue-WPIoT system. 
	
	
	To solve the resource allocation and optimization, we put forth an optimization framework with the objective of minimizing the joint energy-and-latency cost during a Ue-WPIoT communication period. In this optimization problem, the energy cost is represented by the supplied energy emitted from the UAV, and the latency cost is essentially the overall time spending on a Ue-WPIoT communication process, depending on the summation of powering time and data transmission time (\ie, Steps \scalebox{1.03}{\textcolor{mygreen}{\ding{184}}} and \scalebox{1.03}{\textcolor{blue}{\ding{185}}}). It is worth mentioning that the UAV may consume a large portion of the hovering propulsion energy during the Ue-WPIoT communication process. The optimization is subject to the constraints including the \textit{sufficient harvested energy} to ensure the successful data transmission, the \textit{limited latency} to ensure the communication task being completed within a given time, and also the feasible values for system parameters (\eg, the WPT power). It is non-trivial to solve this optimization problem due to the multiple factors, convexity and dynamicity of the Ue-WPIoT communication. The solutions to this optimization problem may depend on several optimal variables, such as BF vector, powering duration time, the data transmission power emitted from IoT nodes. 

	Moreover, this optimization framework shall also consider multiple concurrent communications. In particular, the optimal resource allocation scheme needs to be adjusted to power multiple activated nodes which can then transmit the data back to the UAV. Regarding the concurrent multi-node powering strategy, the energy BF vector at the UAV needs to be optimized, thereby maximizing the received power at multiple nodes. In the data transmission procedure, the UAV then needs to conduct a fine-grained task scheduling to fulfill the multi-access data transmission. Particularly, time-division-multiple-address (TDMA), space-division multiple access (SDMA) and orthogonal frequency-division multiple access (OFDMA) schemes can be adopted to support multiple accesses~\cite{PLi:TWC20}. In TDMA, each IoT node will be assigned with a time slot, during which the data can be transmitted to the UAV. Consequently, the data transmission from multiple activated IoT nodes will be scheduled one after the other. The SDMA scheme leverages the BF technology to spatially separate the multiple data transmissions. Specifically, the UAV can adjust a dedicated BF vector for each IoT node through the position estimation of IoT nodes from the feedback signals in the wake-up procedure (\ie, Step \scalebox{1.03}{\textcolor{mymag}{\ding{183}}}).
	

	\subsection{UAV Trajectory Optimization}
	\label{subsec: trajedesign}
	The UAV trajectory in Ue-WPIoT needs to be carefully designed with consideration of multiple factors together. Different from the UAV trajectory design in existing studies, which merely consider the shortest-path solutions to cover all the IoT nodes~\cite{DYang:TVT18}, Ue-WPIoT needs to a joint consideration of multi-node access, multi-task scheduling and energy efficiency optimization. First, the UAV trajectory should be as short as as possible to save the total flight time in Ue-WPIoT. Moreover, multiple nodes can be simultaneously visited by the UAV thanks to the multi-node access mechanisms as discussed in Section~\ref{subsec:tech-reallocation}. In this way, the total flight time can be saved.

    The optimal trajectory in Ue-WPIoT can be designed accordingly. In particular, the UAV can activate and power a set of nodes; this set of nodes is called a WPC group. In the optimal trajectory design, the UAV selects one node from a WPC group as a \emph{traversal} point. The UAV can simultaneously power and communicate with other nodes within in the same WPC group. Then, the optimal trajectory becomes the shortest path to visit every traversal point (instead of visiting every IoT node). Intuitively, this trajectory design can achieve a shorter travelling path length than that of a pure shortest-path strategy, which visits every node in one-by-one manner. Ref.~\cite{Zhou:TOC18} presents a similar trajectory design, which also partitions the whole network into several regions. With each region, one node is selected as the visited point in each region though this design only considers geographical partitions of nodes. In contrast, our design optimizes the trajectory with consideration of multi-node access and WPC. 
	
	In our optimal trajectory design, the number of nodes within a WPC group has the impact on the overall travelling path length. For example, more nodes within a WPC group may lead to a shorter travelling path. The number of nodes within a WPC group depends on the flying height of the UAV. In particular, we denote the flying height of the UAV by $H$, the maximum powering distance by $d_{EH}$, which is essentially the distance between the UAV and the IoT node. We then project the powering distance to the 2D plane in order to calculate the coverage area in the plane. The horizontal coverage range (\ie, the projection $d_{EH}$ on the plane) is denoted by a ground circle with radius $R$, which needs to fulfill the condition $R=\sqrt{d_{EH}^2-H^2}$ according to the triangular relation. When the WPT power is fixed (\ie, $d_{EH}$ is fixed), the lower flying height of the UAV leads to more IoT nodes to be covered, implying a shorter travelling path. The numerical results as shown in Section~\ref{subsec: trajedesign} will further confirm this observation.
	
	

	\section{Numerical results and analysis}	
	\label{sec:simu}
	
	This section provides numerical results to demonstrate the effectiveness of enabling technologies mainly from the following two aspects.
	
	
	\subsection{Achievable EH distance and Data rate}
	\label{subsec:simuwpc}

	We first present simulation results on the achievable EH distance and the data transmission rate in Ue-WPIoT. We consider that an IoT node can harvest the wireless energy emitted by one UAV and consequently use the harvested energy to transmit the sensory data to the UAV. It is worth mentioning that the achievable EH distance (\ie, the maximum WET range) can be derived by analyzing the radio propagation condition between the UAV and one IoT node. The data rate can be obtained by solving the optimization problem of minimizing overall energy consumption of the UAV for WET. In particular, the achievable data rate is derived after substituting the optimal data transmission power into Shannon capacity, where the optimal data transmission power power is mainly determined by the harvested power at the IoT node.
	

	In our simulations, the A2G channel model is determined by the probability-based Line of Sight (LoS)/Non Line of Sight (NLoS) path loss under suburban geographical parameters. It is worth mentioning that the LoS component usually dominates the A2G channel when the UAV is in the high altitude~\cite{Foto:CST19} while the NLoS component may have higher impact than the LoS component especially in the low altitude, where there are obstacles such as trees and bushes~\cite{YShi:WCNC18}. Thus, the probability-based LoS and NLoS model can take both the cases into account, especially considering the remote or rural area. Moreover, the energy transmitting power at the UAV is fixed at $10\mathrm{W}$. The energy conversion efficiency at the IoT node is set as $0.3$. The communication bandwidth for data transmission is $15\mathrm{MHz}$. We choose different AE-BF settings (as discussed in Section~\ref{subsec:AE-BF}) to evaluate the achievable powering distance and data rate.

	\begin{enumerate}
		\item We consider that a UAV is equipped with an omnidirectional antenna or a BF antenna consisting of multiple antenna elements. In particular, we denote the number of antenna elements by $N$, which is equal to 1, 16, and 32, where 1 antenna element denotes an omnidirectional antenna (or an isotropic antenna); $N=16$ and $N=32$ represent 16 antenna elements and 32 antenna elements, respectively. Generally, the more antenna elements imply the higher antenna gain (the more directivity toward a direction). 
		\item We adopt three carrier frequencies, 400MHz, 900MHz, and 2.4GHz for WPT. Note that each IoT receiver (\ie, energy harvester) has its input power threshold corresponding to the different carrier frequencies. In particular, as indicated in the latest study~\cite{hak2019lowpowersensing}, the 400MHz-based energy harvester requires the input power at least $-20\mathrm{dBm}$, while the 900MHz-based energy harvester and the 2.4GHz-based energy harvester can support the input power threshold of $-23\mathrm{dBm}$ and $-50\mathrm{dBm}$, respectively.
	\end{enumerate} 

	\begin{figure}[t]
		\footnotesize
		\centering
		\subfloat[Under carrier frequency 400MHz]{\includegraphics[width=6cm]{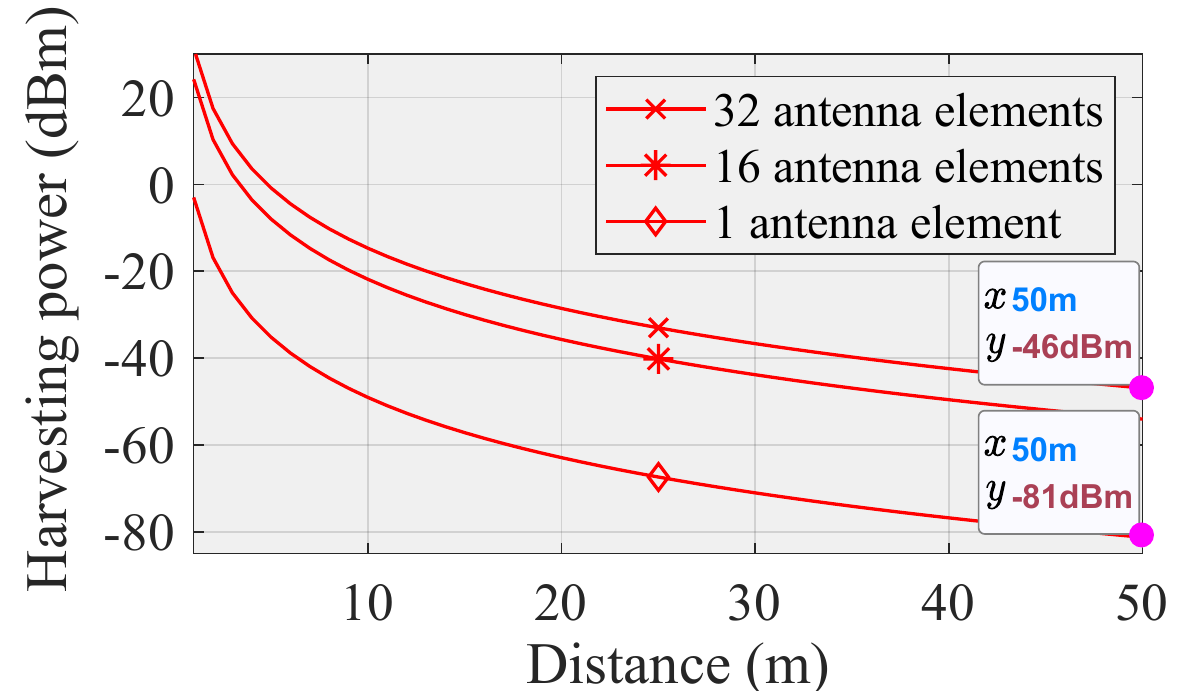}\label{fig:ehcompare1}}\hfil
		\subfloat[Under 32 antenna elements]{\includegraphics[width=6cm]{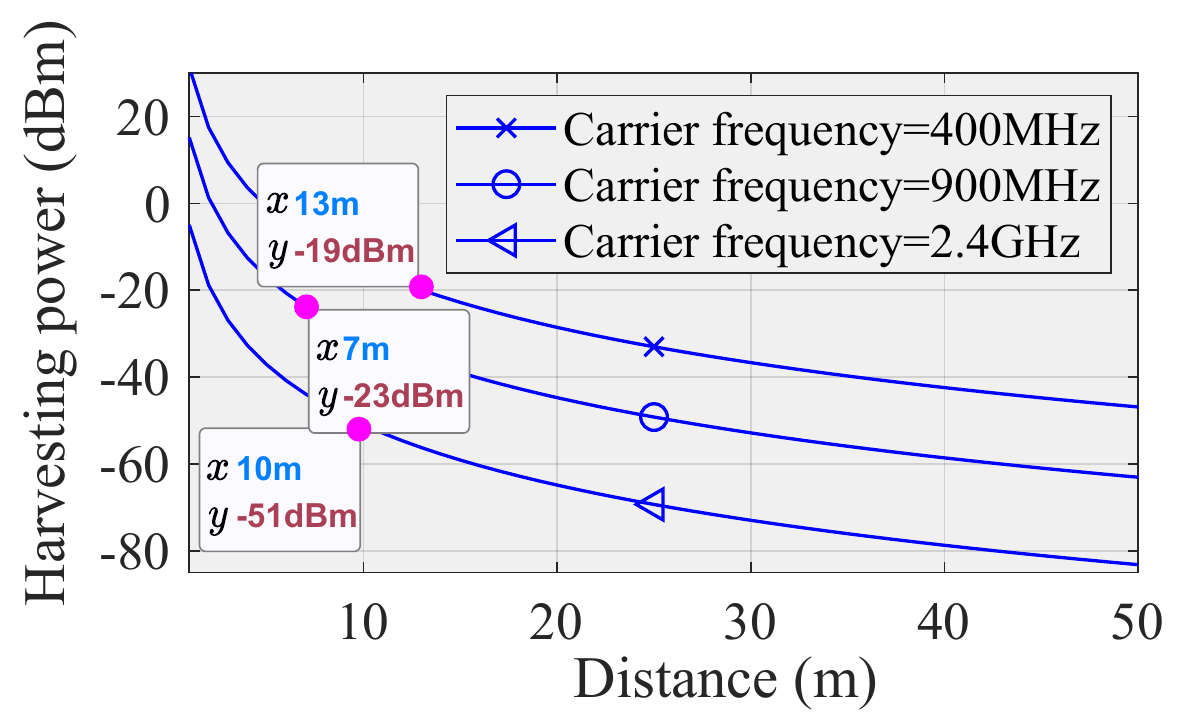}\label{fig:ehcompare2}}\hfil	
		\caption{Harvested power at the IoT node versus the distance, carrier frequency and number of antenna elements, with the energy transmitting power being $10\mathrm{W}$ and the energy conversion efficiency being $0.3$.}
		\label{fig:Eharvesting}
	\end{figure}
	
	The above BF antenna can be realized by a uniform planar array (UPA)~\cite{WYi:TC20}. For instance, a 32-element UPA in $4\times8$ element-array with 400MHz carrier frequency has an approximate size about $1.125\times2.625\mathrm{m}^2$. Equivalently, 32-elements UPA antennas with 900MHz carrier frequency and 2.4GHz have approximate sizes about $0.5\times1.16\mathrm{m}^2$ and $0.1875\times0.4375\mathrm{m}^2$, respectively. Large size UAVs such as MQ-1 Predator drones (used for US Army) and civilian drones like Zipline drones\footnote{\url{https://spectrum.ieee.org/robotics/drones/in-the-air-with-ziplines-medical-delivery-drones}.} can be potentially adopted for WPC tasks in the future.


	Fig.~\ref{fig:Eharvesting} presents the harvested power at IoT node versus the distance with varied carrier frequencies and different numbers of antenna elements. In particular, Fig.~\ref{fig:ehcompare1} plots the harvested power at IoT node versus the distance when the carrier frequency is fixed at 400MHz. We observe from Fig.~\ref{fig:ehcompare1} that the harvested power drops dramatically with the increased distance due to the path loss effect over the long distance. Moreover, Fig.~\ref{fig:ehcompare1} also shows that the increased number of antenna elements can counteract the path loss effect. For example, when the number of antenna elements is 32, the harvested power at $10\mathrm{m}$ is still above $-20\mathrm{dBm}$ (\ie, the threshold of input circuit at the harvester) while omnidirectional antenna does not reach this threshold. This is because the more antenna elements implies the higher antenna gain (\ie, the more directivity of an antenna). 
	
	Moreover, the carrier frequency also affects the harvested power. Fig.~\ref{fig:ehcompare2} plots the harvested power at IoT node versus the distance when the number of antenna elements is fixed at 32. We observe from Fig.~\ref{fig:ehcompare2} that the lower frequency can lead to a longer achievable energy harvested range. For example, the achievable EH distance is $13\mathrm{m}$ when the carrier frequency is 400MHz and the minimum input power $-19\mathrm{m}$. 

	\begin{figure}[t]
		\footnotesize
		\centering
		\subfloat[Under carrier frequency 400MHz]{\includegraphics[width=5.5cm]{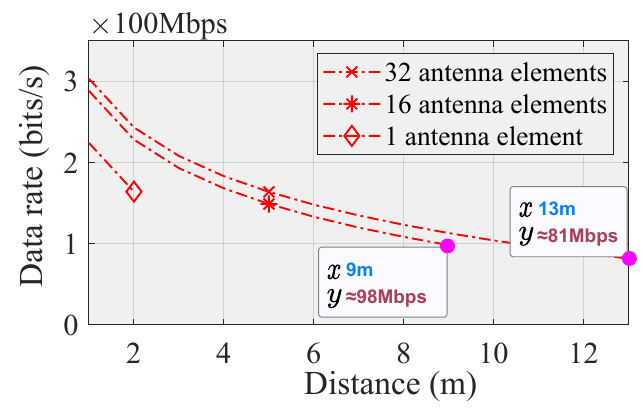}\label{fig:rcompare1}}\hfil
		\subfloat[Under 32 antenna elements]{\includegraphics[width=5.5cm]{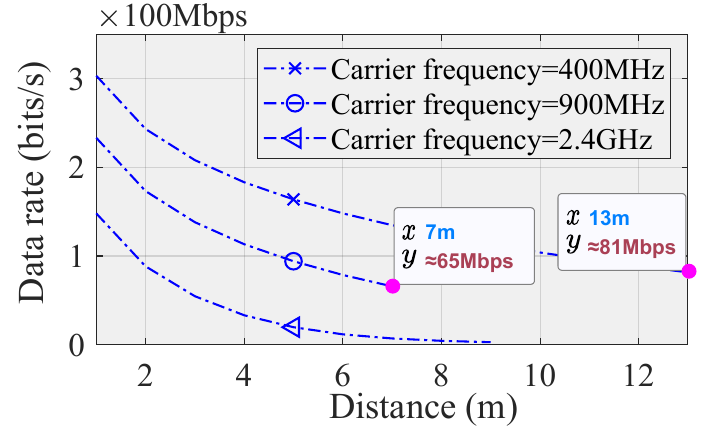}\label{fig:rcompare2}}\hfil
		\caption{Achievable data rate versus the distance, carrier frequency and antenna elements, with the bandwidth being 15MHz.}
		\label{fig:dtransmitting}
	\end{figure}

	Fig.~\ref{fig:dtransmitting} presents the achievable data rate versus the distance with varied carrier frequencies and different numbers of antenna elements. Obviously, the achievable data rate has a similar trend to the harvested power. In particular, Fig.~\ref{fig:rcompare1} plots the achievable data rate when the carrier frequency is fixed at 400MHz. We also find that the achievable data transmission rate decreases with the increased distance due the path loss effect while the more antenna elements can compensate for the path loss. Fig.~\ref{fig:rcompare2} plots the achievable data rate versus the distance when the number of the antenna elements is fixed to 32. We have similar findings to the harvested power, that is, the lower carrier frequency can compensate for the data rate loss. It is worth mentioning that data rates in an order of magnitude above 50Mbps are achievable in most of settings as shown in Fig.~\ref{fig:dtransmitting}. For instance, the data rate of $65\mathrm{Mbps}$ is achieved under the setting of $900\mathrm{MHz}$ and 32 elements as shown in Fig.~\ref{fig:rcompare2}. It implies that Ue-WPIoT can potentially support high data-rate applications.
	
     \begin{figure*}[t]
			\subfloat[Shortest Path in one-by-one manner]{\includegraphics[width=4.4cm]{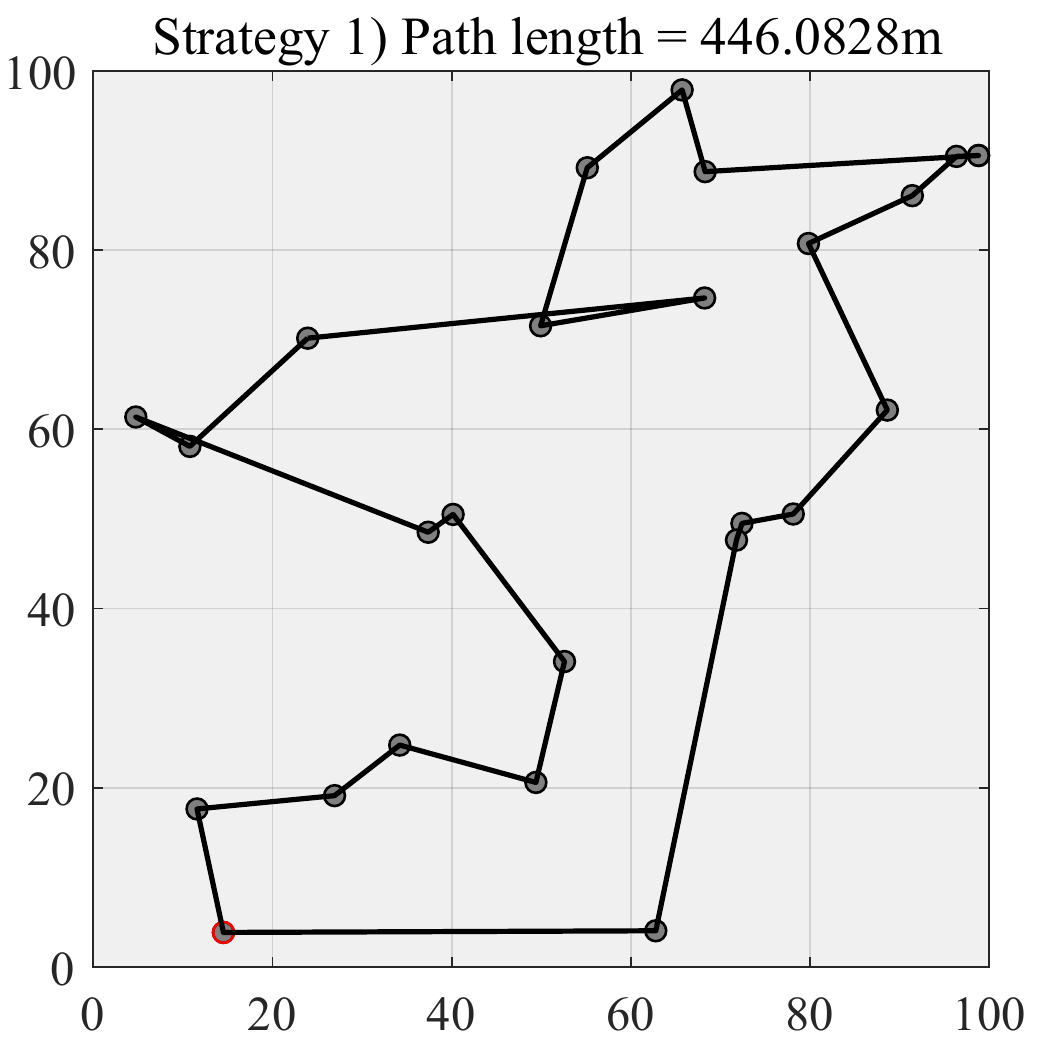}\label{fig:path0}}\hfil
			\subfloat[Shortest Path with multi-node communications with the UAV Height $H=10\text{m},5\text{m}$, respectively]{\includegraphics[width=12.5cm]{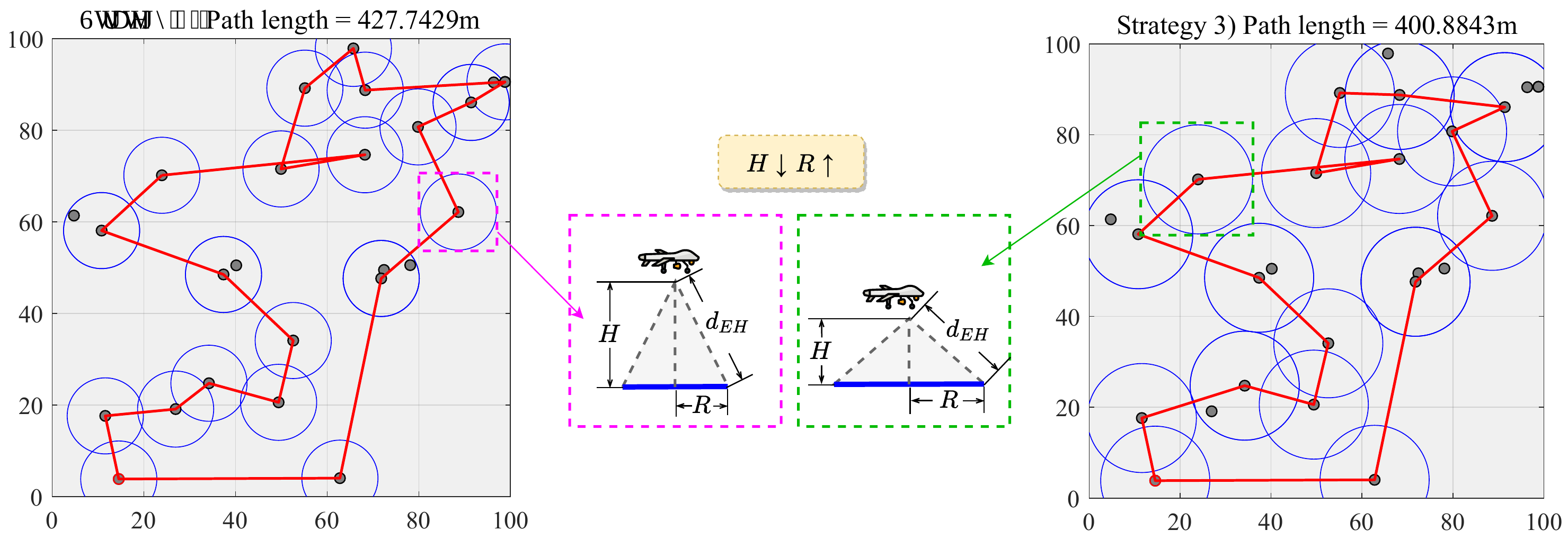}\label{fig:path10}}\hfil
			\caption{The UAV trajectory in the $100\text{m}\times100\text{m}$ area with the nodes density being $0.25$, where the blue circle denotes the achievable powering range.}
			\label{fig:UAVpath}
	\end{figure*} 

	\subsection{Trajectory Design for multi-node communications}
	\label{subsec:simutrajectory}
	We next conduct simulations to analyze the UAV trajectory with consideration of multi-node communications. In particular, our experiments were conducted in a $100\mathrm{m}\times 100\mathrm{m}$ area where the IoT nodes are randomly distributed with density of $0.25$, as shown in Fig.~\ref{fig:UAVpath}. Implied by the results in Section~\ref{subsec:simuwpc}, we consider the following optimal settings: 1) the maximum EH distance being given by $d_{EH}=13\mathrm{m}$, 2) the antenna consisting of 32 antenna elements, 3) the carrier frequency being fixed at 400MHz. In addition, to ensure a feasible powering range, the UAV's flying height $H$ requires to fulfill the triangular relation as shown in Fig.~\ref{fig:path10}. 
	
	In our simulation, we consider three trajectory strategies: 1) the shortest-path trajectory when the UAV covers every IoT node in one-by-one manner; 2 the designed shortest-path trajectory with multi-node communications when the UAV hovers at height being $10\textrm{m}$ to serve multiple IoT nodes within the same achievable EH distance $d_{EH}$; 3) the designed shortest-path trajectory with multi-node communications when the UAV hovers at height being $5\textrm{m}$ to serve multiple IoT nodes within the same achievable EH distance $d_{EH}$. Regarding strategies 2) and 3), the horizontal coverage range of the UAV can be derived by the triangular relation. Specifically, given $H=10\mathrm{m},5\mathrm{m}$ in strategies 2) and 3), the corresponding horizontal coverage circles have the radius $R_2=\sqrt{13^2-10^2}\approx8.31\mathrm{m}$ and $R_3=\sqrt{13^2-5^2}=12\mathrm{m}$, respectively. We observe that strategy 3 has a larger coverage area than strategy 2, implying more IoT nodes potentially falling into a WPC group.
	

	
	Fig.~\ref{fig:UAVpath} compares three trajectory strategies. 
	It is observed that strategies 2) and 3) achieve the travelling path length values with $427.7429\mathrm{m}$ and $400.8843\mathrm{m}$, respectively when the flying height is $H=10\mathrm{m},5\mathrm{m}$, respectively. In contrast, the travelling path length in strategy 1) is $446.0828\mathrm{m}$, which is much longer than those in strategies 2) and 3). This is because our trajectory design with consideration of multi-node communications can serve for multiple nodes within one WPC group while the conventional one-by-one shortest path strategy can only cover one node at a time. In addition, we can observe that the travelling path length of strategy 3) is even shorter than that of strategy 2). The reason may lie in more nodes to be served in strategy 3) when the horizontal coverage range $R=12\mathrm{m}$, which is longer than $R\approx8.31\mathrm{m}$ of strategy 2), thereby further shortening the travelling path. 
	 
	Although a lower flying height can lead to a larger coverage area, the low-flying UAV may be susceptible to obstacles (such as trees or rocks) at a low altitude. The recent advances in autonomous UAV manipulation and auto-navigation bring the opportunities to address this issue~\cite{ELBANHAWI201727}. For example, Artificial Intelligence (AI) assisted computer vision technologies can help UAVs detect obstacles.  
	
	\section{Conclusion and future directions}\label{sec:conc}
	
	In this article, we present an overview of UAV-enabled wireless powering Internet of Things (Ue-WPIoT), which can potentially overcome two major constraints of IoT: 1) energy constraint of IoT nodes and 2) difficulty in deploying and maintaining infrastructure nodes in remote or rural area. We then elaborate the design challenges of Ue-WPIoT and discuss the enabling technologies to address these challenges. Simulation results validate the effectiveness of the presented solutions. We outline several future directions in Ue-WPIoT as follows.
	
	\subsection{Resource limitation of UAVs}
	
	In Ue-WPIoT, UAVs are serving as both energy suppliers and data collectors. Although the optimization of wireless energy transferring and data collection processes can somehow save energy of UAVs, UAVs still suffer from a substantial energy consumption. Energy charging or fuel filling may severely affect the trajectory and the coverage of UAVs, especially in the remote and rural area. Thus, energy-harvesting UAVs from ambience will be an important future direction. The possible energy-harvesting technologies for UAVs include energy harvesting from solar panels and the adoption of windmilling propellers.
	
	\subsection{Trajectory Privacy Protection}
	
	The trajectory information of UAVs is a prerequisite for the effective control, efficient route planning and navigation, especially in adverse weathers or disaster situations. However, UAVs can be vulnerable to malicious attacks such as hijacking after stealing or intercepting the trajectory information~\cite{wu2019safeguarding}. For example, UAVs can be tracked, intercepted and even hijacked once the trajectories of UAVs are exposed to malicious users~\cite{XSun:WM19}. Moreover, the behaviours of UAV users can be tracked and deduced through analysing the trajectories of UAVs. Therefore, the trajectory privacy protection of UAVs will be an important future direction.
	
	\subsection{Intelligent Algorithms for Trajectory Design of UAVs}
	As analyzed in the earlier part of this article, the trajectory of UAVs in Ue-WPIoT needs to consider multiple factors, such as multi-node communications and priority of data collection tasks. However, it is challenging to design optimal trajectories for UAVs with consideration of all these factors together. In addition, the dynamicity of IoT (\eg, the failure of some IoT nodes) makes the situation even worse, \ie, the pre-designed trajectory needs to be adjusted. The advent of AI, deep learning, reinforcement learning brings the opportunities to address this rising challenge. The lightweight or portable AI models are expected to be designed for UAVs in the future. 

	\balance
	\bibliographystyle{IEEEtran}
	\bibliography{ref}
	
	\begin{IEEEbiographynophoto}{Yalin Liu} is pursuing her Ph.D. degree in Faculty of Information Technology at Macau University of Science and Technology. Her current research interests include Internet of Things and Unmanned aerial vehicle communications. 
	\end{IEEEbiographynophoto}
	
	\begin{IEEEbiographynophoto}{Qubeijian Wang} is currently with in School of Cyber Security, Northwestern Polytechnical University, Taicang, China. His current research interests include Internet of Things, wireless security and Unmanned aerial vehicle communications. 
	\end{IEEEbiographynophoto}
	
	\begin{IEEEbiographynophoto}{Hong-Ning Dai} [SM’16] is currently with the Department of Computing and Decision Sciences, Lingnan University, Hong Kong, as an associate professor. He obtained the Ph.D. degree in Computer Science and Engineering from Department of Computer Science and Engineering at the Chinese University of Hong Kong. His current research interests include the Internet of Things and blockchain technology. He has served as associate editors of IEEE Transactions on Industrial Informatics, IEEE Systems Journal, and IEEE Access. He is also a senior member of the ACM. 
	\end{IEEEbiographynophoto}
	
	\begin{IEEEbiographynophoto}{Muhammad Imran} is working as a senior lecturer in the School of Engineering, Information Technology \& Physical Sciences, Federation University, Australia. His research interests include mobile and wireless networks, Internet of Things, cloud and edge computing, and information security. He has published more than 200 research articles in reputable international conferences and journals. His research is supported by several grants. He serves as an associate editor for many top ranked international journals. He has received various awards.
	\end{IEEEbiographynophoto}
	
	\begin{IEEEbiographynophoto}{Nadra Guizani} is a Clinical Assistant Professor at the School of Electrical and Computer Engineering at Washington State University. She received her PhD from Purdue University, West Lafayette, USA in 2020. Her research interests include machine learning, mobile networking, big data analysis and prediction techniques. She is an active member of both the Women in Engineering program and the Computing Research Association.
	\end{IEEEbiographynophoto}
	
\end{document}